\documentclass[12pt]{article}
\usepackage[dvips]{graphicx}
\newcommand{\be}{\begin{equation}}
\newcommand{\ba}{\begin{array}}
\newcommand{\bd}{\begin{displaymath}}
\newcommand{\ee}{\end{equation}}
\newcommand{\ea}{\end{array}}
\newcommand{\ed}{\end{displaymath}}
\newcommand{\noi}{\noindent}

\def\lsim{\mathrel{\rlap{\lower4pt\hbox{\hskip1pt$\sim$}}
    \raise1pt\hbox{$<$}}}      
\def\gsim{\mathrel{\rlap{\lower4pt\hbox{\hskip1pt$\sim$}}
    \raise1pt\hbox{$>$}}}      

\def\frac#1#2{{#1\over #2}}
\def\Imag#1{\Im{\rm m}#1}
\def\Real#1{\Re{\rm e}#1}
\def\ds {\displaystyle}

\def\t{$|t|$}
\def\g2{ GeV$^2$}
\def\gm2{ GeV$^{-2}$}

\def\ie{\hbox{\it i.e. }}
\def\etc{\hbox{\it etc... }}
\def\eg{\hbox{\it e.g. }}

\newcommand{\NPB}[1]{Nucl.\ Phys.\ B {\bf #1}}

\newcommand{\PRD}[1]{Phys.\ Rev.\ D {\bf #1}}

\begin{document}

\begin{titlepage}
\title{
\flushright{LYCEN 2001-20}
\bigskip
\bigskip
\bigskip
\bigskip
\begin{center}
{\Large{\bf Rescattering corrections\\
in elastic scattering}}
\end{center}
}
\bigskip
\maketitle
\begin{center}
{\large {
P. Desgrolard $^{a,}$\footnote{E-mail: desgrolard@ipnl.in2p3.fr},
M. Giffon     $^{a,}$\footnote{E-mail: giffon@ipnl.in2p3.fr},
E. Martynov   $^{b,}$\footnote{E-mail: martynov@bitp.kiev.ua}
}}
\end{center}
\bigskip
\bigskip
\noindent
$^a$ Institut de Physique Nucl\'eaire de Lyon, IN2P3-CNRS et
Universit\'e C. Bernard,\\
43 boulevard du 11 novembre 1918, F-69622 Villeurbanne Cedex, France

\noindent $^b$ Bogolyubov Institute for Theoretical Physics, National
Academy of
Sciences of\\
Ukraine, 03143 Kiev-143, Metrologicheskaja 14b, Ukraine

\bigskip
\bigskip
\bigskip
\bigskip
\begin{center}
\begin{minipage}[t]{14.0cm}
{\bf Abstract} A detailed study of the rescattering series is performed
within a model, using a generalized procedure of eikonalization and fitted
to the $pp$ and $\bar pp$ elastic scattering data. We estimate and compare
the various rescattering corrections to be added to the Born contribution
in the amplitude. We find that their number is finite, whereas it
increases with the energy and the transfer, like does their importance. In
the domain where data exist, we find also that a correct computation must
include, at least, all two- and three-Reggeon exchanges and some four- and
five-Reggeon exchanges. Any approximation aiming to reduce this (large)
number of exchange would be hazardous, especially when extrapolating. We
extend our estimates in the domain of future experiments.
\end{minipage}
\end{center}

\bigskip

\end{titlepage}
\section{Introduction}
The different aspects of Reggeon rescattering (or Regge cuts) have been
investigated since the pioneering work of Gribov~\cite{grib}, who first
developed a Regge calculus. A well defined procedure for determining
individual diagrams corresponding to any multi-Reggeon exchanges has been
developed in~\cite{ter}. However these works and the following ones have
been devoted mainly either to the study of analytical forms for the cuts
or to the study of various summation schemes of multi-Reggeon
diagrams~\cite{hist1}. Even when all the rescattering corrections were
taken into account in the fits~\cite{hist2,poes} to experimental data, as
a rule, determining the relative importance of various individual
$n$-Reggeon exchange contributions has not received much interest (see
however in~\cite{dj} where this aspect is discussed).

Motivated by the experiments prospected~\cite{futexp} at RHIC and LHC,
intended to measure the conventional observables in new ranges of energy
and transfer, we are concerned by the following question. How should we
take into account the rescattering corrections to the Born approximation
in a correct computation of those observables~?

To answer this question, one generally uses an eikonalization procedure
which is also a remedy to cure the shortcoming of amplitudes violating the
Froissart-Martin bound~\cite{fm} at the Born level. However such a
procedure is not unique and generally it involves a numerical integration
in the Fourier-Bessel's transform of the eikonalized amplitude over the
impact-parameter ("$b$"). Furthermore the eikonalization, as a global
process, hides the physical origin in terms of "Reggeon"
exchanges~\footnote{ We affect the generic name "Reggeon" to any component
of the elastic scattering amplitude we discuss for $pp$ and $\bar pp$
process \ie Pomeron and Odderon as well as $f$- and $\omega$-subleading
trajectories.}.

Our aim is to investigate numerically the rescattering corrections to the
Born approximation~: their relative importance, their physical meaning in
terms of various Reggeon exchanges and their minimal number required by a
correct reproduction of experimental data. For that purpose, we use a
so-called generalized eikonalization (GE) procedure~\cite{eikge} recently
applied to Regge models (\eg \cite{poes})and fitted to elastic $pp$ and
$\bar pp$ scattering data.

In the amplitude describing the scattering process in the squared
energy-transfer ($s,t$) space, we can
separate the Born contribution and the rescattering series
\be\label{eq1}
\ba{rll}
A^{\bar pp}_{pp}(s,t)\ =& A^{\bar pp}_{pp;GE}(s,t)\\
=&\ {a^{\bar pp}_{pp; Born}}(s,t) \ + {A^{\bar pp}_{pp,rescat}}(s,t)\ ,
\ea \ee with \be\label{eq2} {A^{\bar pp}_{pp,rescat}}(s,t)\ =\
\sum_{n_+=0}^\infty\sum_{n_-=0}^\infty a^{\bar pp}_{pp;n_+,n_-}(s,t)\ .
\ee Each term of this series (\ref{eq2}) is analytically known for the
models under interest. An example of driving the calculations is indicated
in the Appendix~\footnote{The calculation of the non-truncated series is
performed as in~\cite{poes} within the GE procedure~\cite{eikge}.}. We
shall demonstrate that the series is conveniently approximated with a
finite (although not small) number of terms. This possible truncation
allows an easy study of the number and specificity of the exchanges that
we must keep in the infinite summation to obtain a good accuracy in the
final evaluation of the observables~: the total cross-section,
$\sigma_{\rm tot}$, the ratio of the forward real to imaginary parts of
the amplitude, $\rho$, the differential cross section, $\ds{{d\sigma\over
dt}}$. Furthermore, with a suitably chosen Born amplitude, it avoids the
time consuming numerical integration, generally required by any complete
eikonalization procedure.

Each component of the series is labelled with two indexes ($n_\pm$), each
of them having a specific physical meaning. It is straightforward to
constat that the first contribution to the rescattering series (2), with
($n_+,n_-$) = (0,0), is a sum of all diagrams of two-"Reggeon" exchanges.
In fact, this (0,0) term involves ten exchanges so different as
Pomeron-Pomeron, Pomeron-$f$, Pomeron-Odderon, Pomeron-$\omega$, $f$-$f$,
$f$-Odderon, $f$-$\omega$, Odderon-Odderon, Odderon-$\omega$,
$\omega$-$\omega$. It is easy to see that any term, with $n_+ + n_-=N$, is
the sum of all diagrams with $N+2$ Reggeons. We have no theoretical
argument to sort out the magnitude of terms entering in (2), even when $N$
is as small as 1. When $N\gg 1$ many terms are included in the summation,
with alternated signs inducing many cancellations. So, a careful numerical
examination is interesting to find those exchanges which are the most
important.

For the present estimation, we adopt the final amplitude of~\cite{poes},
which corresponds to a so-called "Dipole Pomeron" (DP) model~\footnote{
Actually in this model the Pomeron "dipole", linear combination of a
simple and a double pole in the angular momentum $J-$plane, as explicated
in~\cite{jenko}, is complemented by 2 standard Reggeons, $f$ and $\omega$,
by an Odderon dipole conveniently multiplied by an exponential damping
factor.} with the GE method.

Such a choice has been made because it is a recently published amplitude
respecting the Froissart-Martin bound, implying an involved formalism for
the most general treatment up to now (to our knowledge) of the
eikonalization process including 3 added free parameters. This
complication may of course obscure the physical sense, but the fit is
satisfying for the forward and non-forward data up to the largest explored
$|t|$ (14 \g2 at the ISR, neighboring the Regge limit of application). It
is, on our opinion, necessary to account data in the widest range of high
energies and transfers to get confidence on predictive power outside the
fitted sets of data (remember, the TOTEM project~\cite{futexp} plans
measurements at the Large Hadron Collider up to at least $|t|=10 $ \g2 and
it is precisely the LHC -or Tevatron or RHIC- energy the most interesting
to discuss at present).

The results we found (driven entirely by an analytical calculation) have
not only an illustrative character, we have checked that our conclusions
would be also valid for a "Monopole Pomeron" version, as used
in~\cite{petpro}, with this GE procedure.

\bigskip
\section{Rescatterings and amplitude}
To estimate the rescattering effects, in an absolute manner, we choose to
plot the quantities (appearing basically as the most convenient)
\be\label{eq3} \Real{}\ {\rm and}\ \Imag{\ \left[S_{n_+,n_-}\right]},\quad
S_{n_+,n_-}= {\ a_{pp; Born}(s,t)\ +\ a_{pp;n_+,n_-}(s,t) }\ , \ee for
$t=0$ and for some representative $t$'s. We can easily compare the
rescattering corrections with the results of the computations at the Born
level and with the complete GE. We obtain, for the DP amplitude borrowed
in~\cite{poes}, Figs.~1-6 yielding, on a linear scale~\footnote{ For
commodity in the visualization on a linear scale, the amplitudes with
their original normalization~\cite{poes} have been divided here by $s$.},
the imaginary (left) and real (right) parts of (3), labelled by +($n_+,
n_-$) and plotted versus the energy $\sqrt{s}$, (including the highest
projected LHC energy) for six selected values of $t$ $= 0.,\ -0.5,\ -1.,\
-2.,\ -5.,\ -10.$ \g2.

Alternatively, to estimate the rescattering effects
relatively to the Born result, we may also rewrite the
imaginary part of the amplitude

\begin{figure}[ht]\label{fig.1}
\begin{minipage}[t]{7.0cm}
\begin{center}
\includegraphics*[scale=0.36]{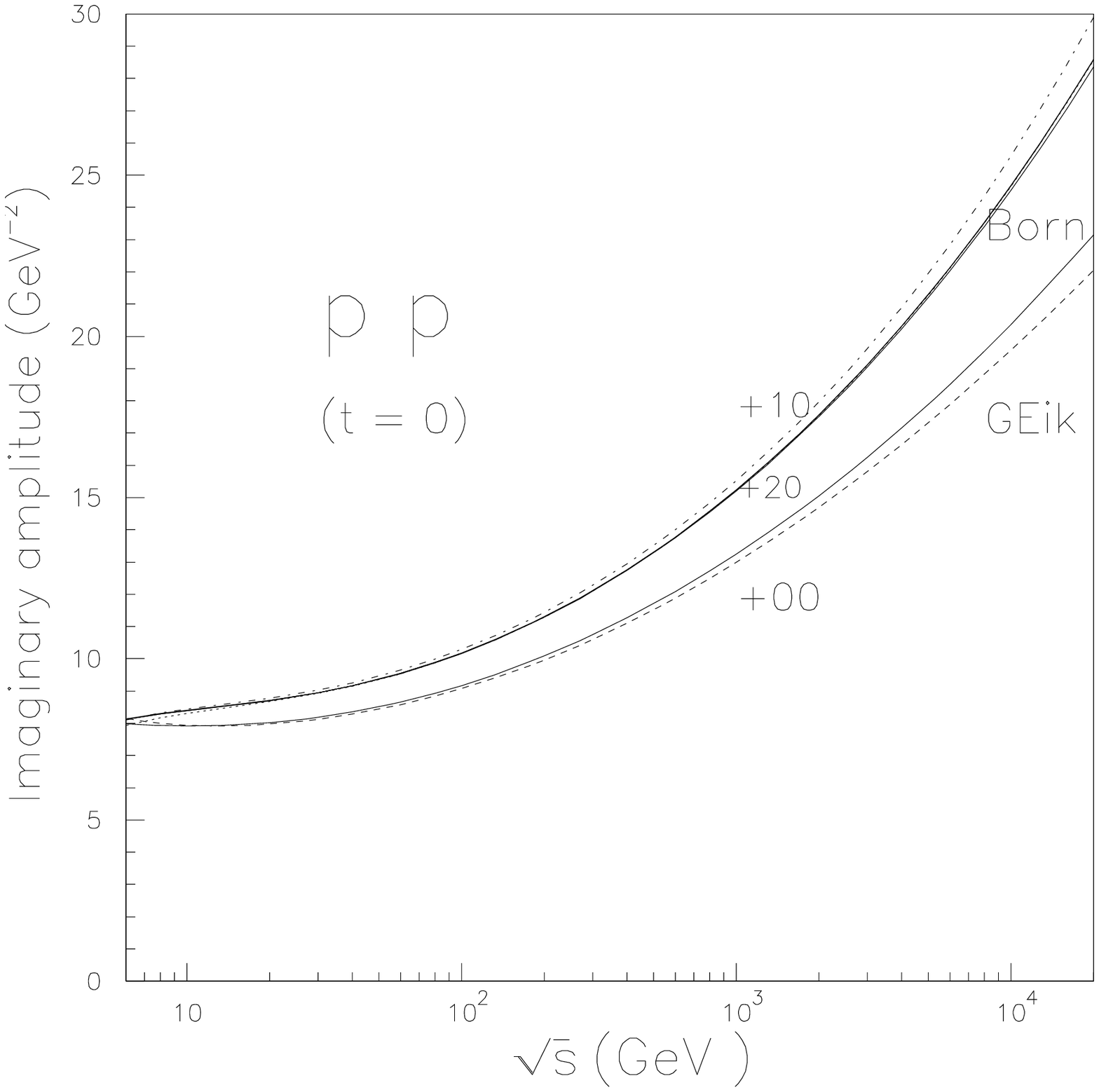}
\end{center}
\end{minipage}
\hskip .8cm
\begin{minipage}[t]{7.0cm}
\begin{center}
\includegraphics*[scale=0.36]{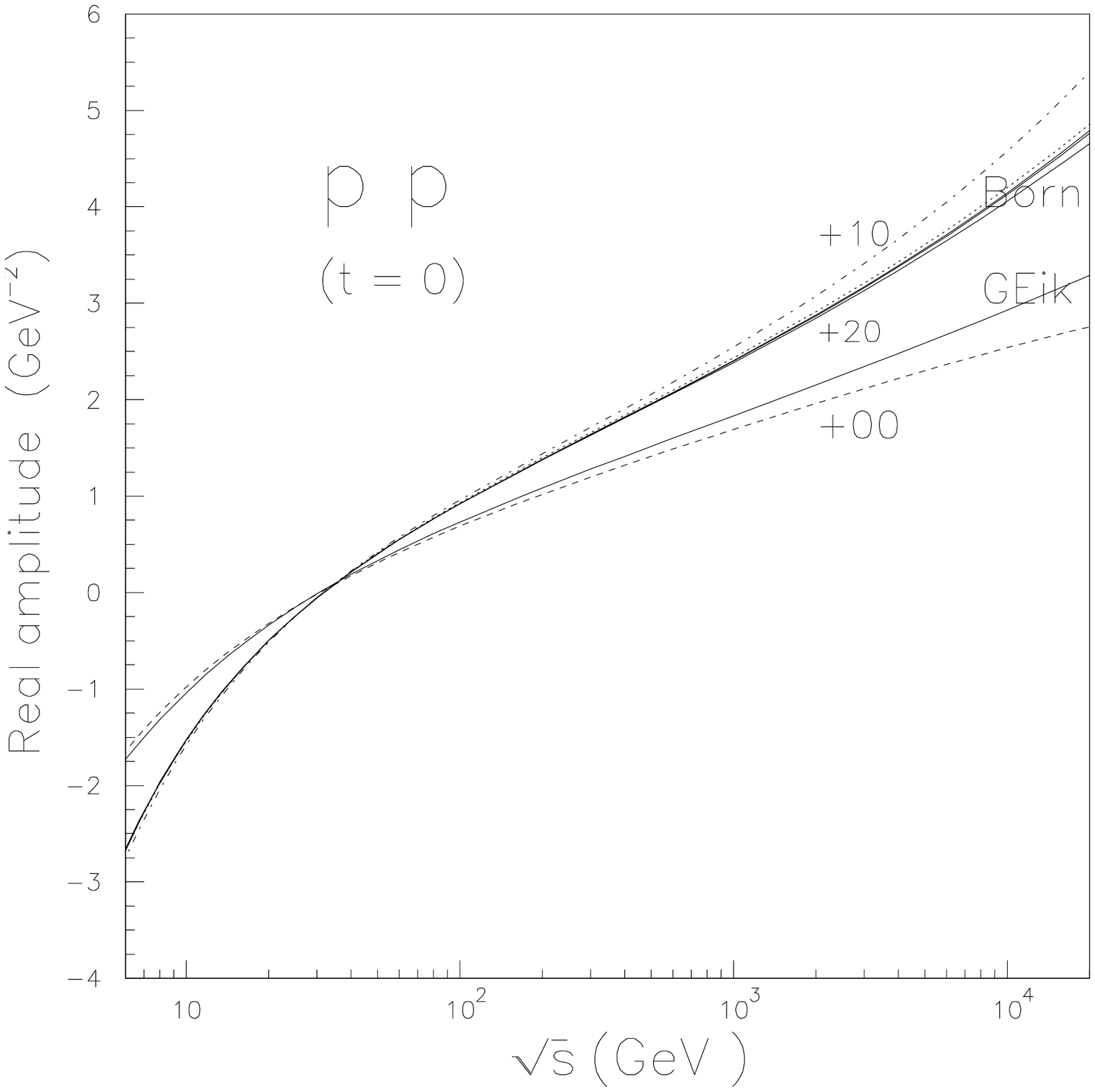}
\end{center}
\end{minipage}
\caption{
 Separate contributions of the main rescattering corrections
(see the text) added to the Born result in the imaginpackfig/ary (left)
and real (right) part of the $t=0$ amplitude ($S_{n_+,n_-}$ (3), in dashed
lines, labelled by $+\ (n_+,n_-)$). Also shown in solid lines are the pure
Born amplitude (Born) and the eikonalized amplitude, once the complete
generalized eikonalization is performed (GEik).}
\end{figure}
\begin{figure}[ht]\label{fig.2}
\begin{minipage}[h]{7.0cm}
\begin{center}
\includegraphics*[scale=0.36]{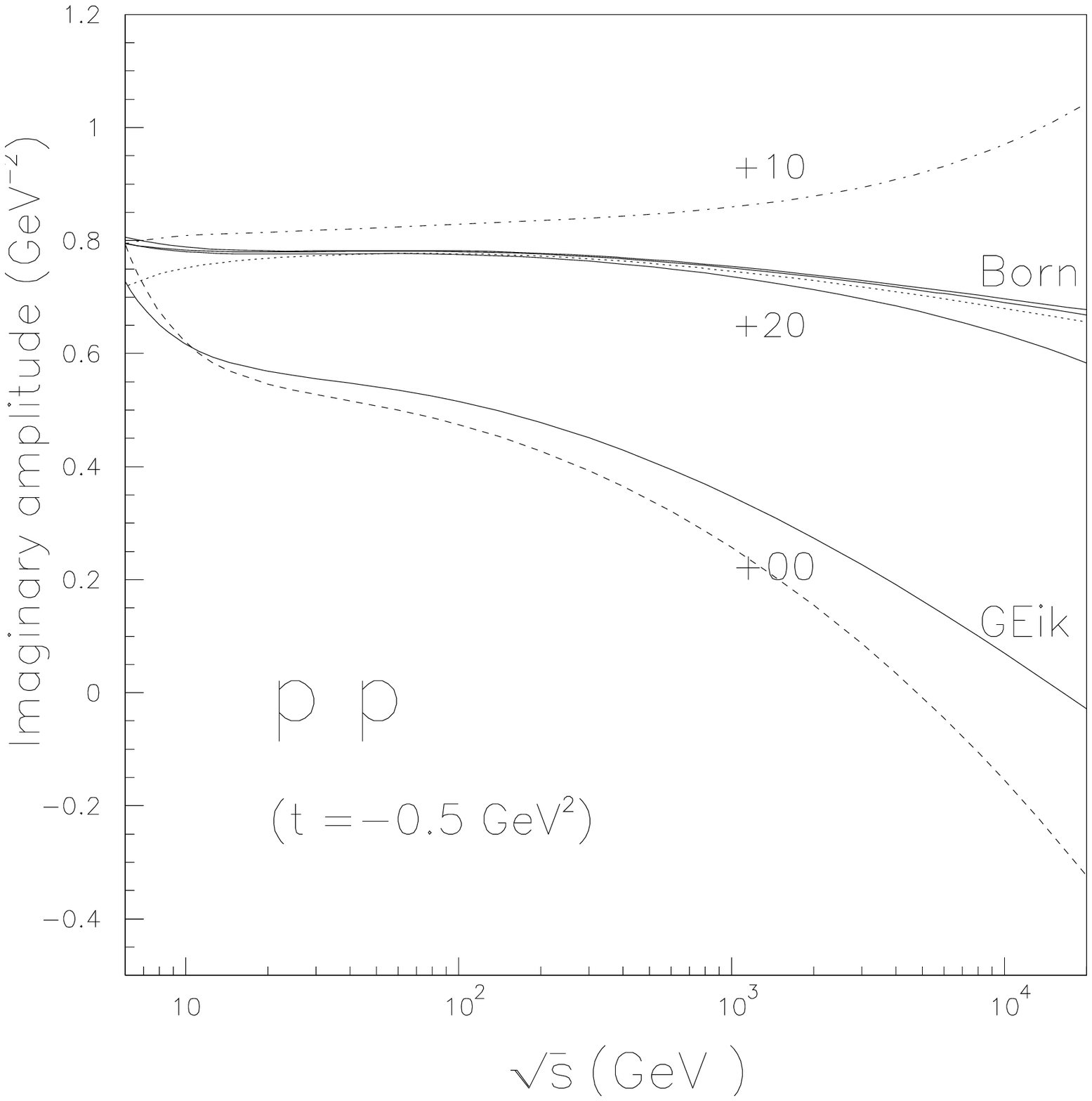}
\end{center}
\end{minipage}
\hskip .8cm
\begin{minipage}[h]{7.0cm}
\begin{center}
\includegraphics*[scale=0.36]{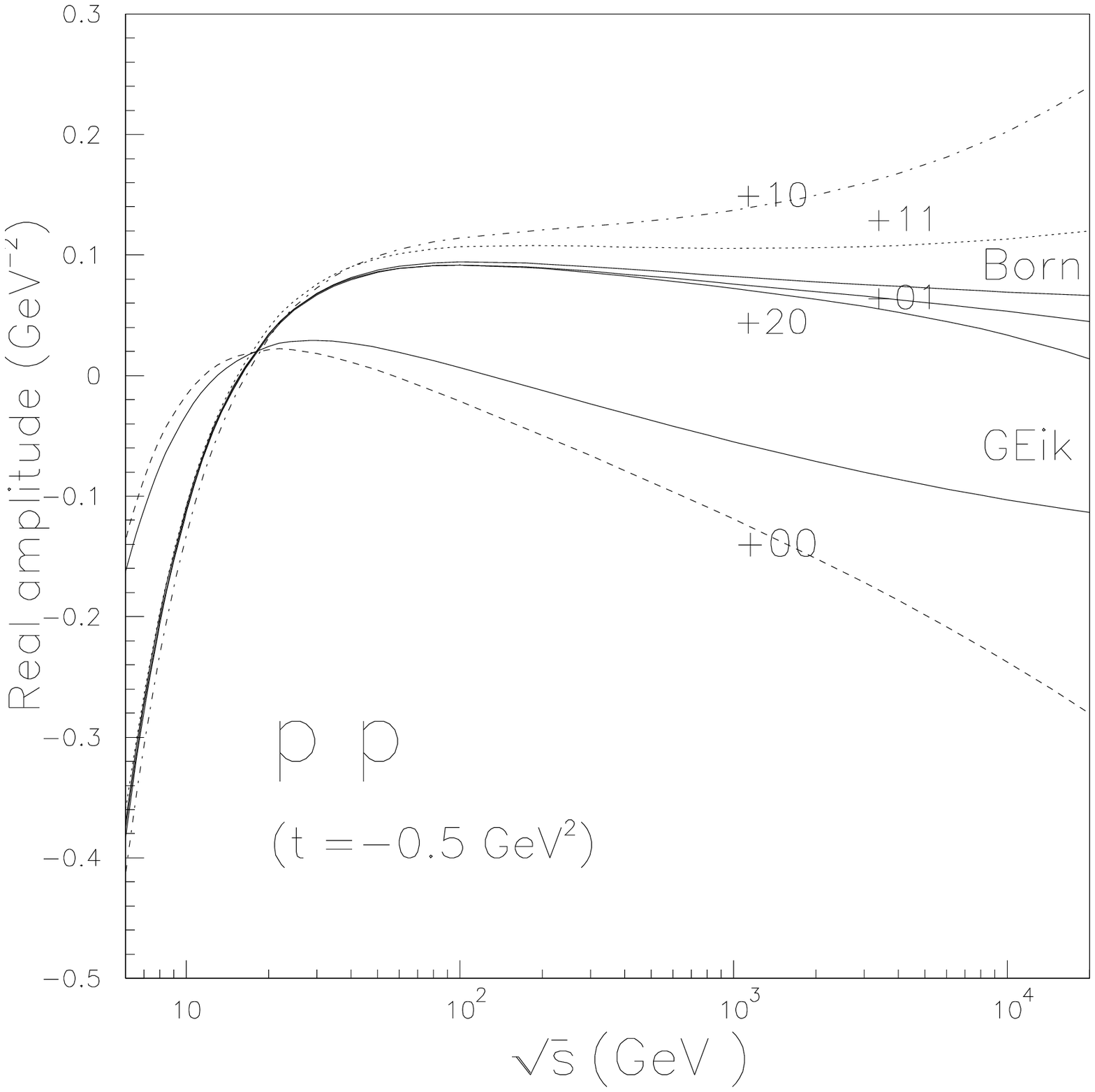}
\end{center}
\end{minipage}
\caption{ Same as in Fig.1 for $t=-0.5$ \g2}
\end{figure}

\be\label{eq4} \Imag{\left(A^{\bar pp}_{pp}(s,t)\right) } = \
\Imag{\left({a^{\bar pp}_{pp; Born}}(s,t)\right)} \ \left( 1 +
\sum_{n_+=0}^\infty\sum_{n_-=0}^\infty R_{n_+,n_-}^{\rm (im)} \right)\ ,
\ee and use this form to settle a hierarchy among the different terms.
Explicitly, \be\label{eq5} R_{n_+,n_-}^{\rm (im)} \ = {
\Imag{\left(a_{pp;n_+,n_-}(s,t)\right)}\over \Imag{ \left(a^{\bar
pp}_{pp,Born}(s,t) \right) } }\ , \ee and similarly for the real part of
the relative rescattering term, defining $R_{n_+,n_-}^{\rm (re)}$.
\begin{figure}[ht]\label{fig.3}
\begin{minipage}[t]{7.3cm}
\begin{center}
\includegraphics*[scale=0.36]{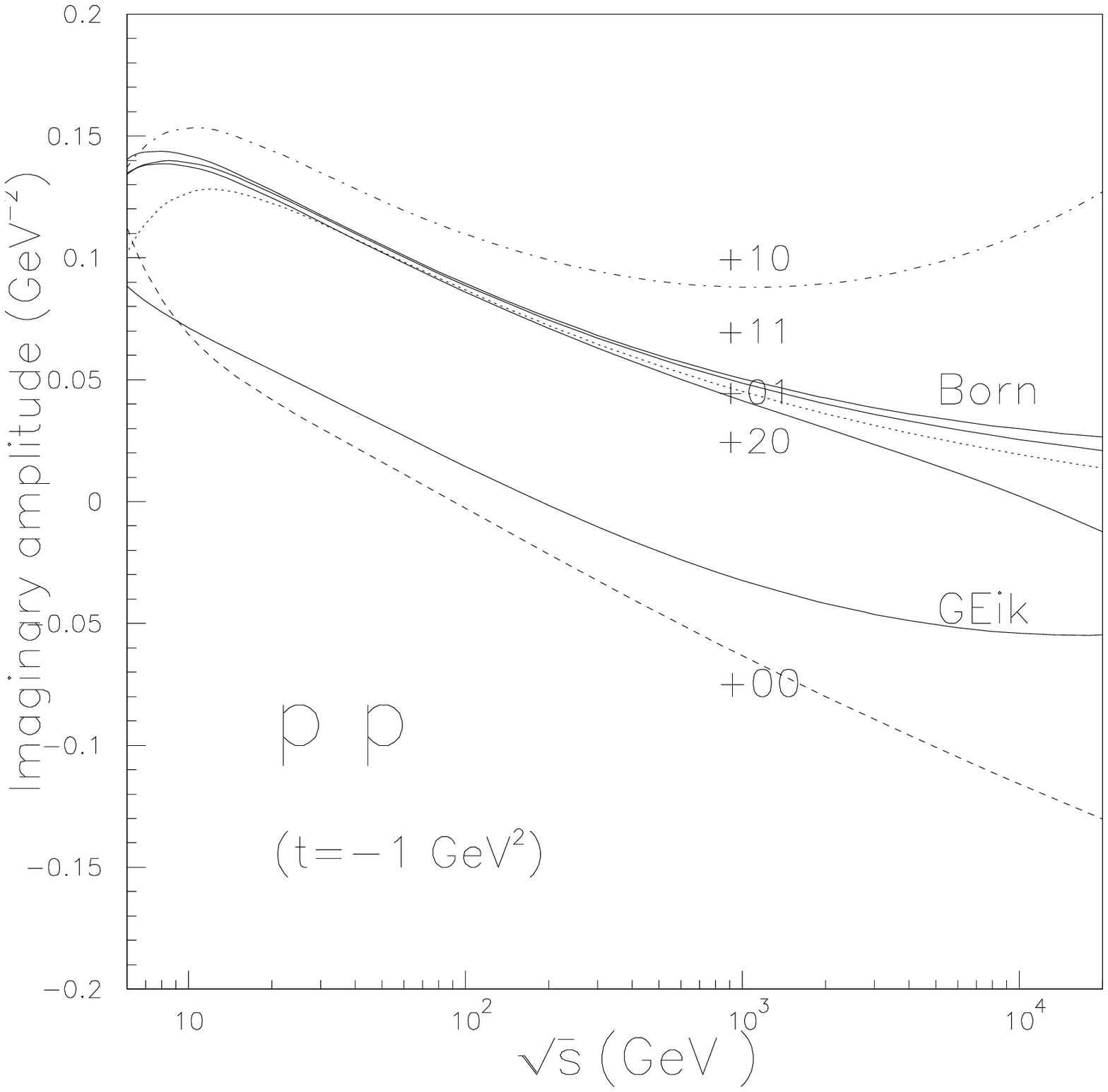}
\end{center}
\end{minipage}
\hskip .2cm
\begin{minipage}[t]{7.3cm}
\begin{center}
\includegraphics*[scale=0.36]{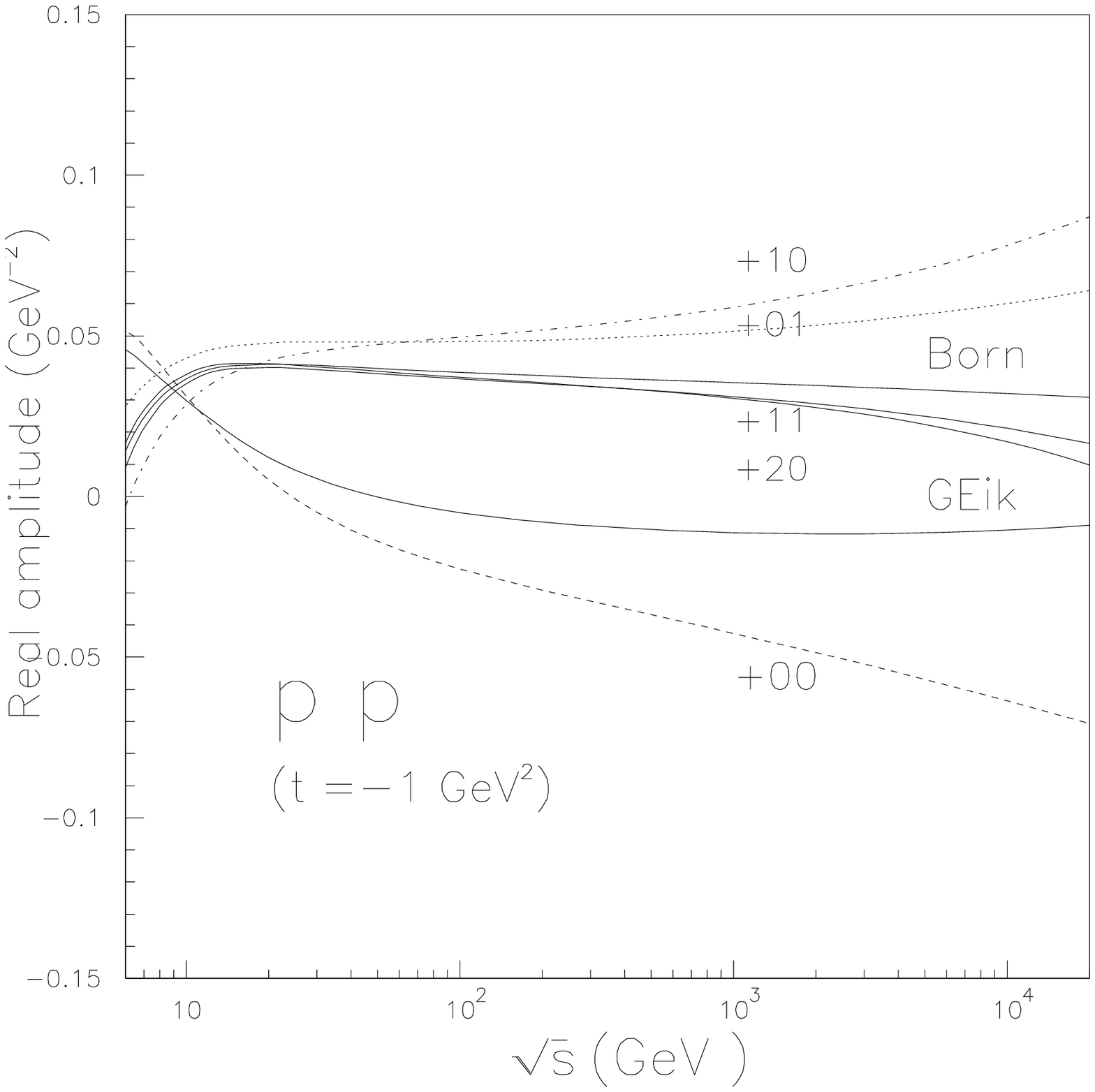}
\end{center}
\end{minipage}
\caption{ Same as in Fig.1 for $t=-1.$ \g2}
\end{figure}
\begin{figure}[ht]\label{fig.4}
\begin{minipage}[t]{7.3cm}
\begin{center}
\includegraphics*[scale=0.36]{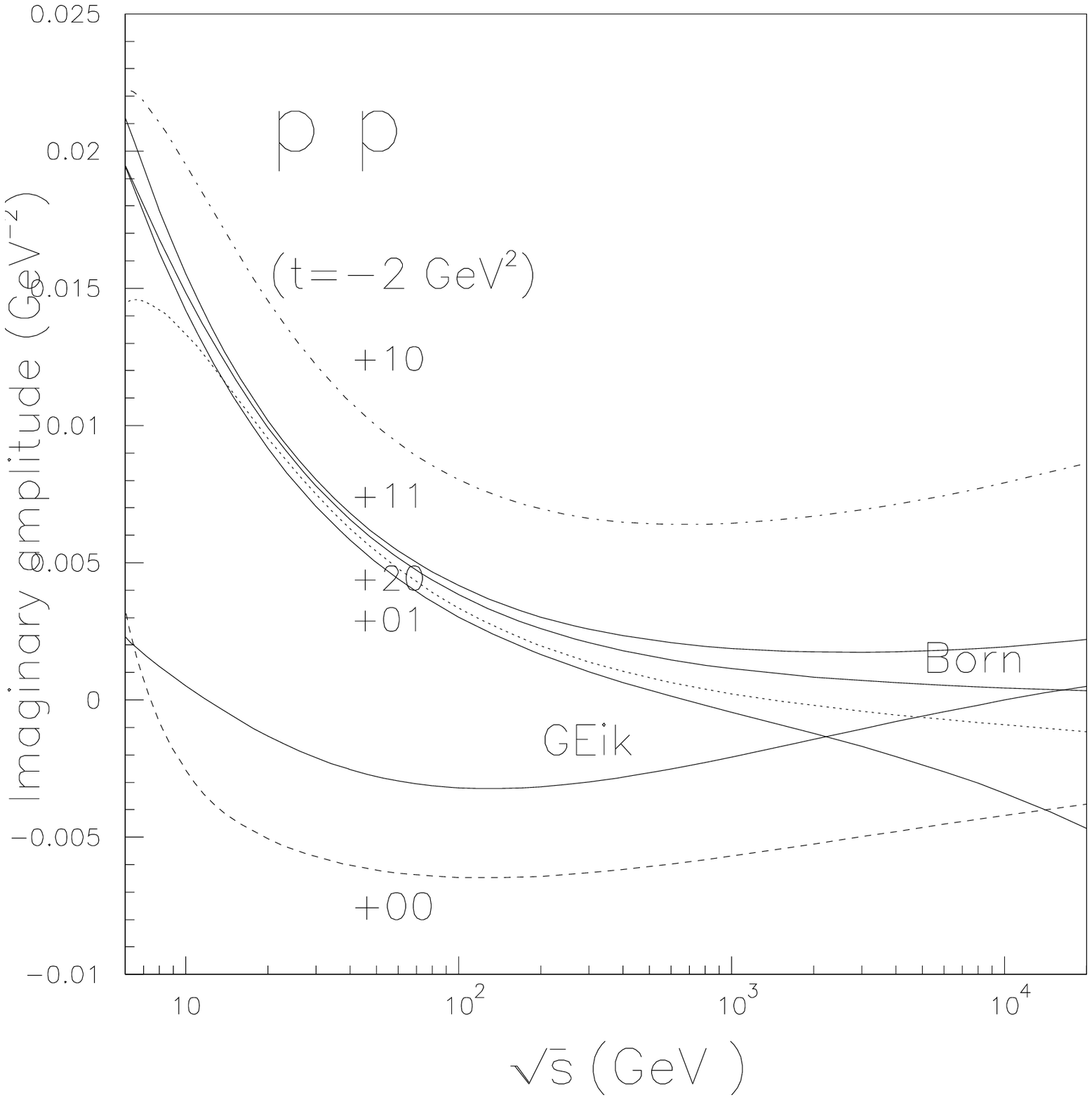}
\end{center}
\end{minipage}
\hskip .2cm
\begin{minipage}[t]{7.3cm}
\begin{center}
\includegraphics*[scale=0.36]{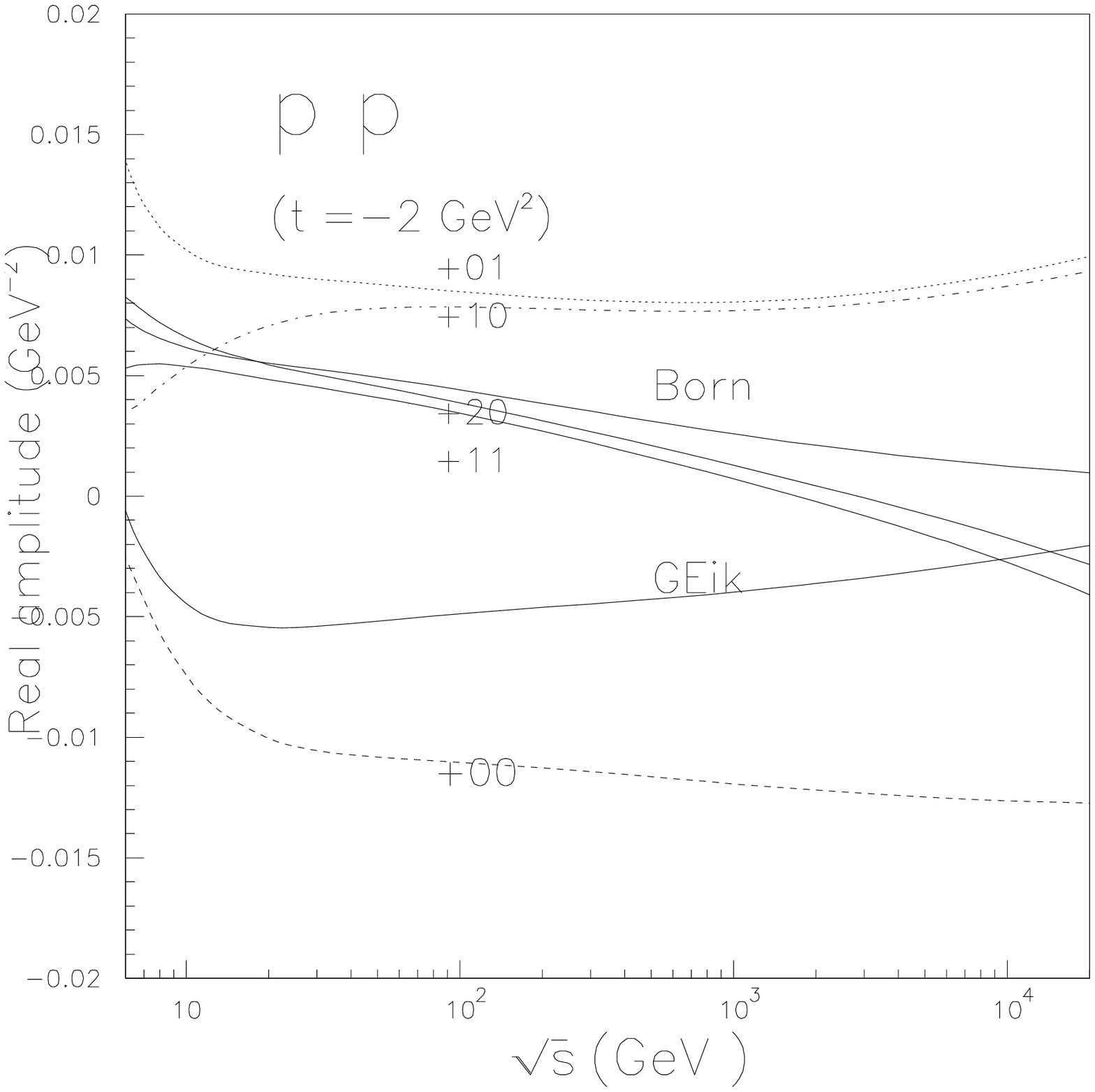}
\end{center}
\end{minipage}
\caption{ Same as in Fig.1 for $t=-2.$ \g2}
\end{figure}
\noi We list in Table 1 an example of these estimations, for the first
values of the indexes $(n_+,n_-)$ (we limit somewhat arbitrarily $R^{\rm
(im)}$ and $R^{\rm (re)}$ to one percent). The examination of the results,
shown in Figs.~1-6 and Table~1, calls for the following comments with
increasing $|t|$~:

 (1) when $t=0$ (see Fig,~1),
adding separate corrections for $n_+>2, n_-> 0$ to the Born term gives
curves that cannot be distinguished by eye from the Born one. In other
terms, a good approximation of the rescattering series is achieved by
keeping only the terms $(n_+, n_-)\ =$ (0,0), (1,0), (2,0). We can see
that the rescattering corrections increase with the energy and we remark
the change of sign and of scale shown by the real part, as required to
account for the experimental characteristics of the $\rho-$ratio.

 (2) when $t$ is in the first diffraction cone (typically $t=-0.5$ \g2, see
Fig.~2), in addition to the three terms already listed for the forward
amplitude, the terms with $(n_+,n_-)=(1,1),(0,1)$ bring small
contributions to the amplitude which increase with the energy like the
three other ones. Like in the forward case, the real and imaginary parts
of the eikonalized amplitude are below the corresponding Born ones.
\begin{figure}[ht]\label{fig.5}
\begin{minipage}[t]{7.3cm}
\begin{center}
\includegraphics*[scale=0.36]{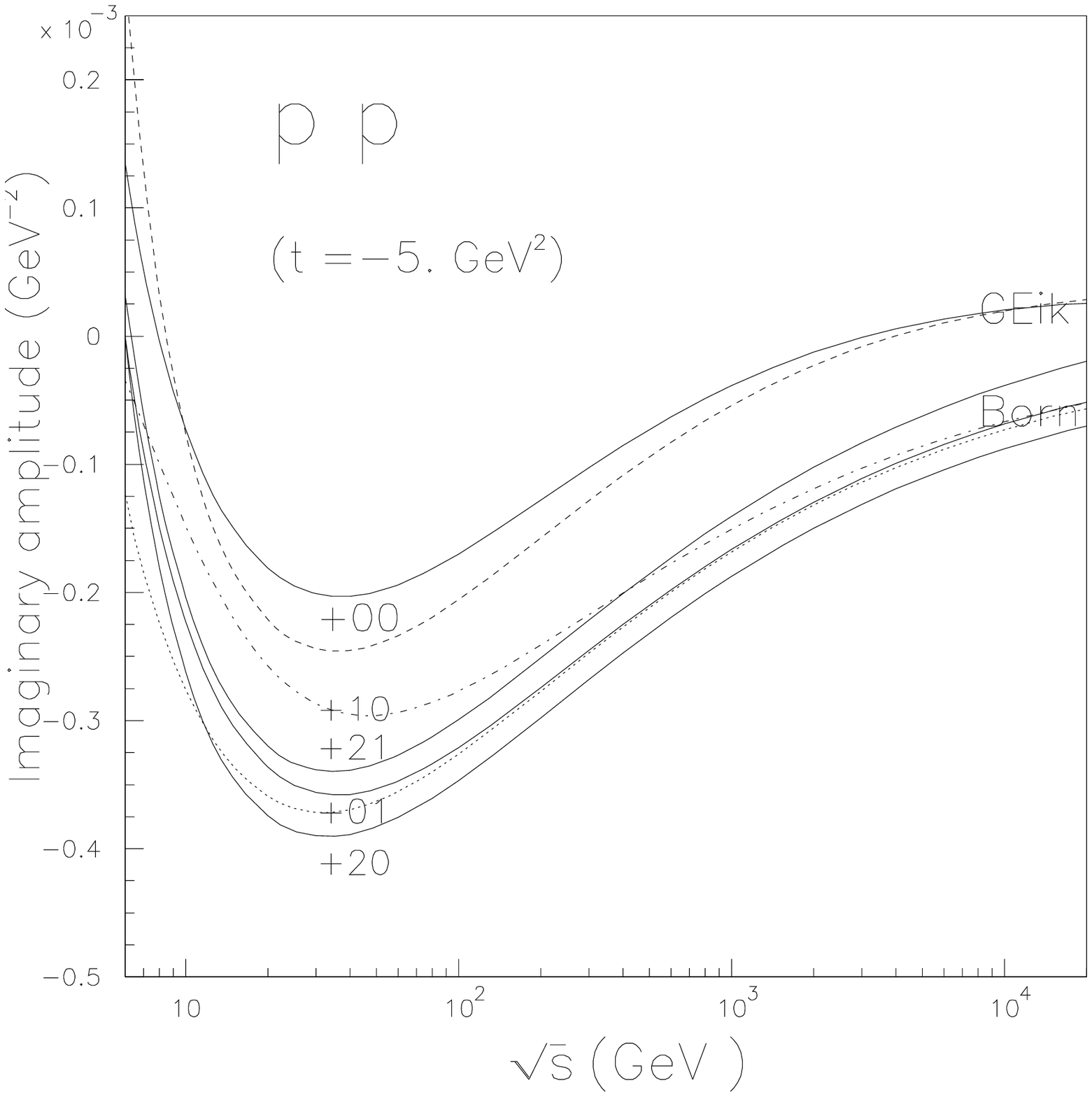}
\end{center}
\end{minipage}
\hskip .2cm
\begin{minipage}[t]{7.3cm}
\begin{center}
\includegraphics*[scale=0.36]{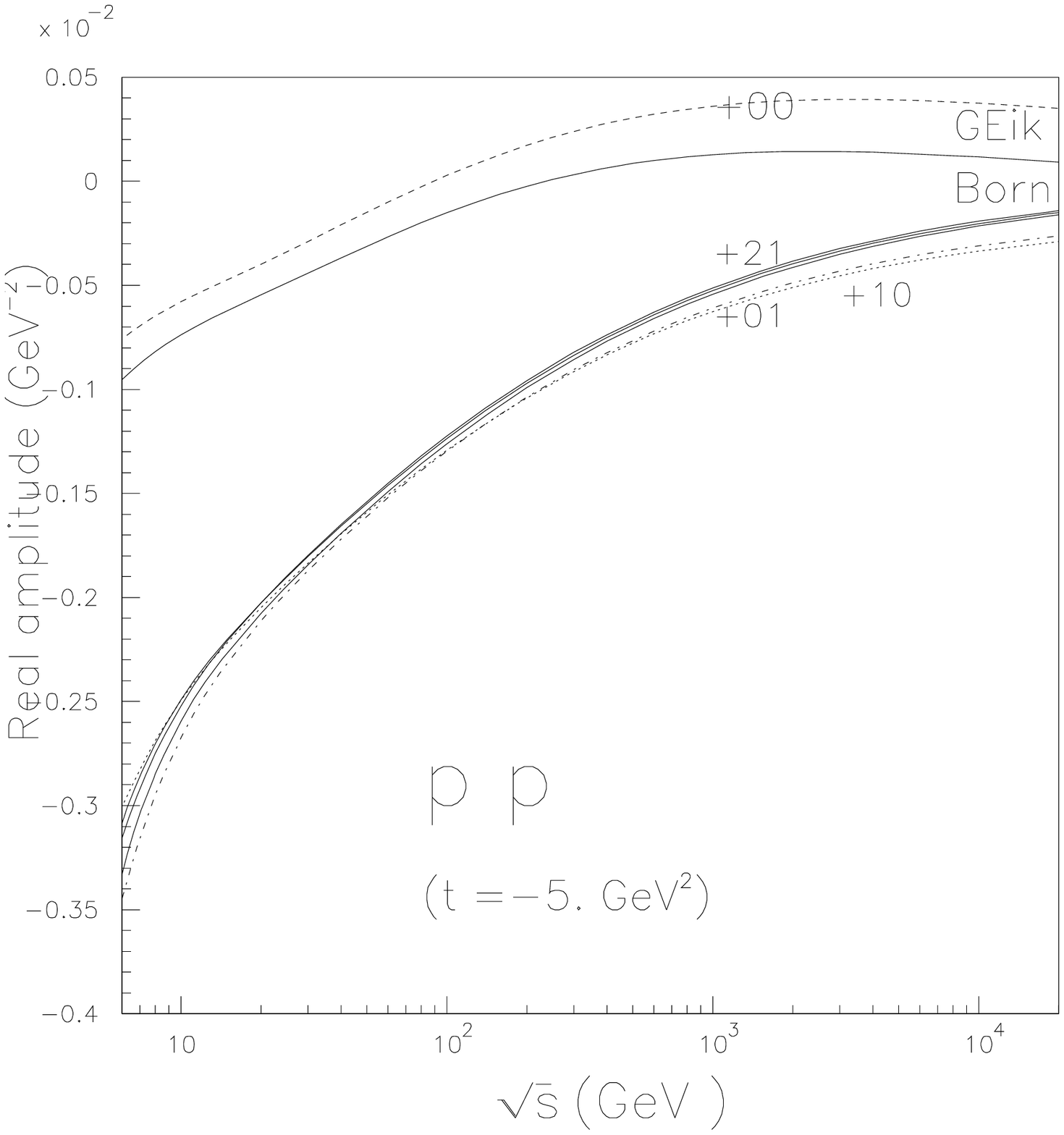}
\end{center}
\end{minipage}
\caption{ Same as in Fig.1 for $t=-5.$ \g2}
\end{figure}
\begin{figure}[ht]\label{fig.6}
\begin{minipage}[t]{7.0cm}
\begin{center}
\includegraphics*[scale=0.36]{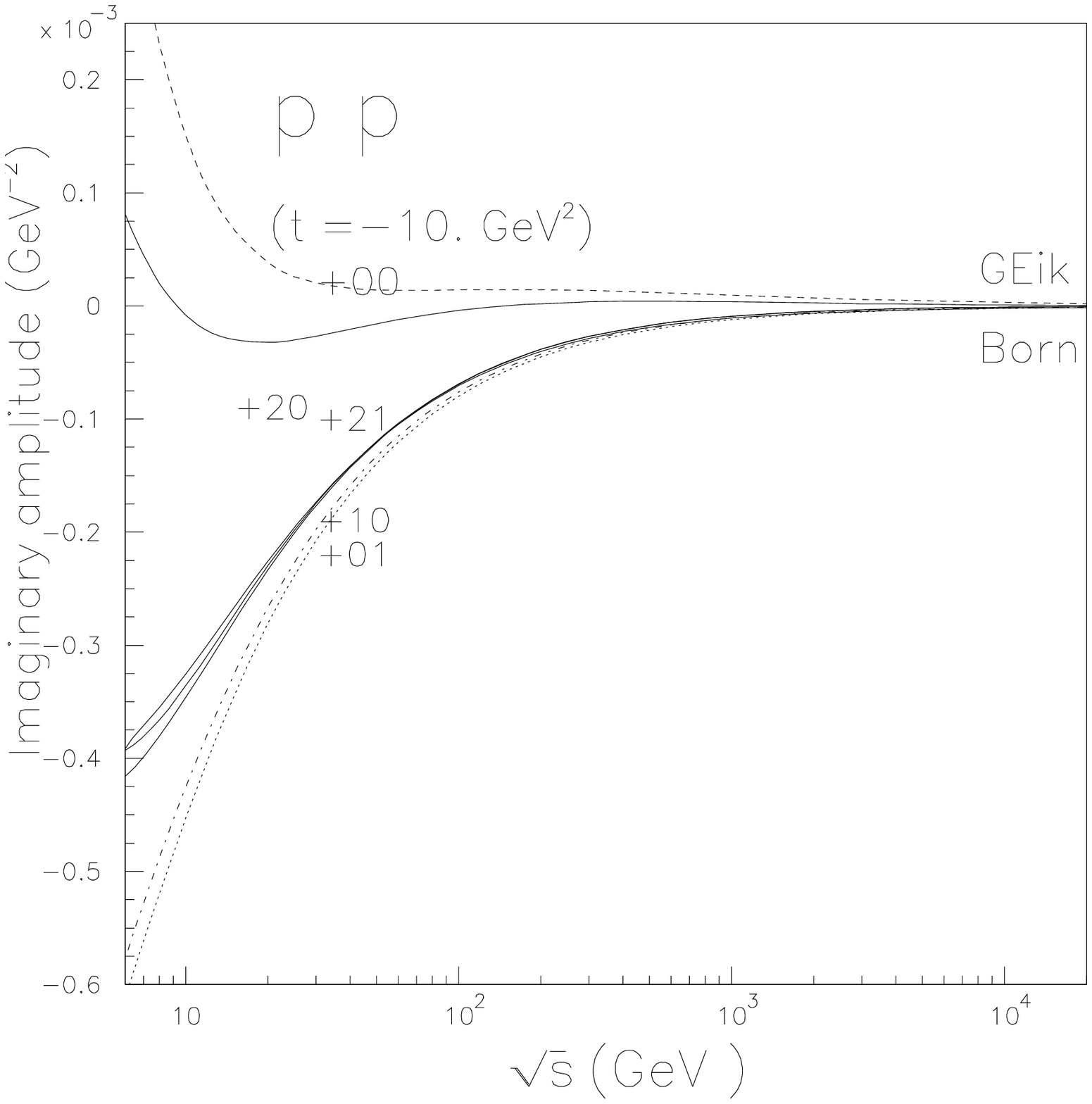}
\end{center}
\end{minipage}
\hskip .8cm
\begin{minipage}[t]{7.0cm}
\begin{center}
\includegraphics*[scale=0.36]{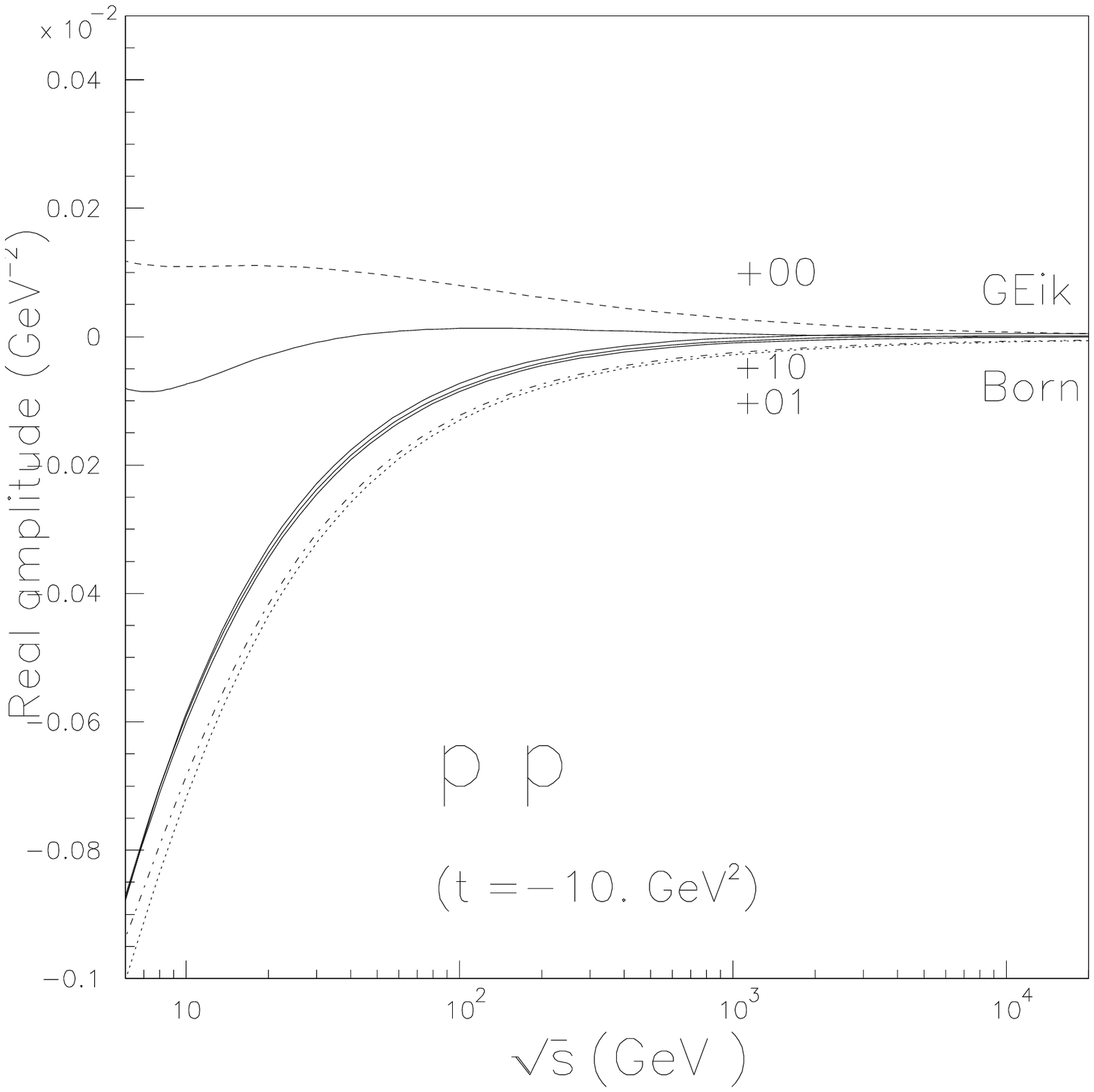}
\end{center}
\end{minipage}
\caption{ Same as in Fig.1 for $t=-10.$ \g2}
\end{figure}

(3) when $t=-1$ \g2 (see Fig.~3) in the vicinity of the dip seen in the
$pp$ angular distributions at the ISR, then, taken one by one, the
corrections due to the rescattering still increase smoothly with the
energy. But, and this is rather surprising, their sum tends towards a
constant value, resulting in a GE limit almost parallel (and below) the
Born estimation as soon as the energy exceeds 50 GeV.

 (4) the value $t=-2$\g2 (see Fig.~4) is in a region where strong
interferences between the various terms exist; however, one remarks
like for $t=-1$ \g2, an individual growth of the corrections with
the energy  and changes of sign resulting in a crossing of the Born and GE
limit; otherwise stated, there is a
clear global compensation of the corrections at the LHC energy for
that  transfer.

 (5) when $t$ has an intermediate value, beyond the first dip-bump
structure ($t=-5$ \g2, see Fig.~5), in addition to the five values of
$(n_+,n_-)$ necessary to have a good precision when estimating the
amplitude, one should also consider the term (2,1). In contrast with the
preceding cases, {\it(i)} both imaginary and real part of the eikonalized
amplitude are above the Born ones {\it (ii)} the growth with the energy of
the absolute values of the corrections due to the rescattering which
saturates at smaller $|t|$ begins to decrease in the considered energy
range (a reminiscence of the dip~?). Finally, we remark that at the
highest investigated energy, the eikonalized amplitude (real and imaginary
part) is almost zero, but the corrections, though very small, may exceed
the Born result by several orders of magnitude. They cannot be neglected
since they bring the main contribution to the angular distributions.

 (6) when $-t$=10 \g2 (see Fig.~6), highest value in the future
prospects~\cite{futexp}, the various corrections tend asymptotically to
become very small at high energy (at the Tevatron and LHC) resulting in a
numerical coincidence between the Born and the GE results corresponding to
a very small differential cross-section. The same remarks as in the
preceding case are valid.
\begin{center}
\begin{tabular}[ht]{|c|c|ccc|}
\hline
$(n_+n_-)$&       &Born         & correction &GE      \\
          &       &      (\gm2) &    (\%)    &(\gm2)  \\
    \hline
      & Imag  & \ 0.752 \   &    \       &\ 0.347 \\
          & Real  & \ 0.082 \   &    \       &$-$0.055\\
\hline
\hline
(00) & Imag &  &$-$65.   & \\
     & Real &  &$-$244.  & \\
\hline
(10) & Imag &  &+14.  & \\
     & Real &  &+66. & \\
\hline
(20) & Imag &  &$-$2.  &\\
     & Real &  &$-$12. & \\
\hline
(01) & Imag &  &$-$  & \\
     & Real &  & +28.  & \\
\hline
(11) & Imag &  &$-$ &\\
     & Real &  &7.  & \\
\hline
(21),(30) & Imag & &$-$ & \\
          & Real & & 1.   &\\
\hline
\end{tabular}
\end{center}
\noindent Table 1. Typical contributions of the main rescattering terms
specified by $(n_+n_-)$ relative to the Born contributions (see the text).
An arbitrary criterion of 1 \% has been set to limit their number. The
energy and transfer are 1000 GeV, $-0.5$ \g2 respectively. Also quoted are
the values of the amplitude computed at the Born level and once the
complete generalized eikonalization is performed (GE).

\smallskip
In summary, it appears that the rescattering corrections to the amplitude
cannot be neglected especially at high energy and transfer. In addition,
the hierarchy of the corrections also depends (as expected) somewhat on
the energy and transfer. As a general rule, for the Regge model considered
here, one can only be sure that $(n_+,n_-)$ = (0,0) brings always the most
important contribution and that limiting these indexes by $n_+=2$ and
$n_-=1$ is probably sufficient at not too high energy and transfer.

\bigskip
\section{Discussion and conclusion}
The total cross-section is directly related to the imaginary part of the
forward amplitude, hence a part of the conclusions of the preceding
section is fully usable. For the $\rho-$ratio, it is less evident to
discuss the effects of the rescattering using the preceding considerations
on the complex amplitude. The differential cross-section being
proportional to the squared modulus of the amplitude, it is not
straightforward to visualize the effects on its behavior when adding
separately the rescatterings. Its non-linear character obscures the
effects we want to investigate. The complete reconstruction of the
amplitude requires the knowledge of both the imaginary and real part for
all ($s,t$) inside the experimental ranges (which is not possible from the
data) and it is not quite evident that the conclusions of the preceding
section on the complex amplitudes still hold for the real observables.

As in~\cite{poes}, when fitting the Dipole or the
Monopole Pomeron model with the GE procedure, in the present work,
many tests have proven than the $\chi^2$ strictly does
not change if we overpass $n_+=2$ and $n_+=1$, keeping more than
$3\times 2=6$ terms in the series (2)
(\ie $(n_-,+n_-)=$ (0,0), (1,0), (2,0), (0,1), (1,1), (2,1) ).
Further, a poor approximation is
realized with two (0,0) and (1,0), or better with three of them (0,0),
(1,0), (2,0).
That does NOT mean that, if we scrutinize in particular some differential
cross-sections at intermediate \t-values
and at high energy, one knows which kind of approximation is to be used,
 because
{\it (i)} the involved data are scarce and then they are depreciated in
the global minimization procedure {\it (ii)} the addition of higher terms
required in the amplitude may have significant consequences on the angular
distribution for some $(s,t)$, meriting a separate study.

\smallskip
The last, but not the least important item we discuss is the use of various
approximations.

A mean to see in which kinematical range an approximation is efficient is
to examine numerically its consequences on the agreement with fitted
observables or, outside the experimentally investigated domain, with the
predictions of non-truncated series (GE case), which reproduces the data.
We choose the $pp$ angular distributions, known at the ISR up to rather
large transfers, and extrapolate up to future experimental conditions. We
comment our results on the following points.

\begin{figure}[ht]\label{fig.7}
\begin{center}
\includegraphics*[scale=0.5]{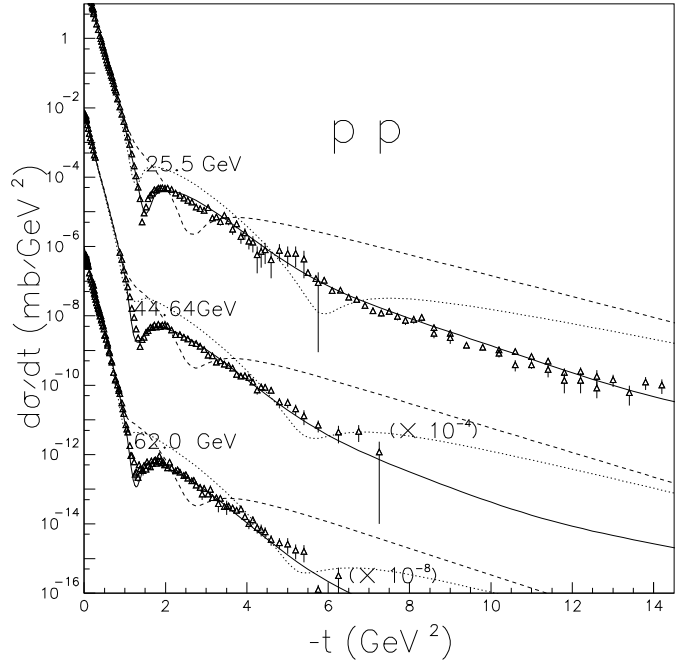}
\end{center}
\vskip -0.5cm \caption{ Modifications of a part of the fit in~\cite{poes}
(solid line) when the rescattering series is approximated with the
two-Pomeron exchanges exclusively (dashed line) or with the all ten
possible two-Reggeon exchanges, \ie $(n_+,n_-)=(0,0)$ (dotted line). }
\end{figure}
\begin{figure}[ht]\label{fig.8}
\begin{center}
\includegraphics*[scale=0.5]{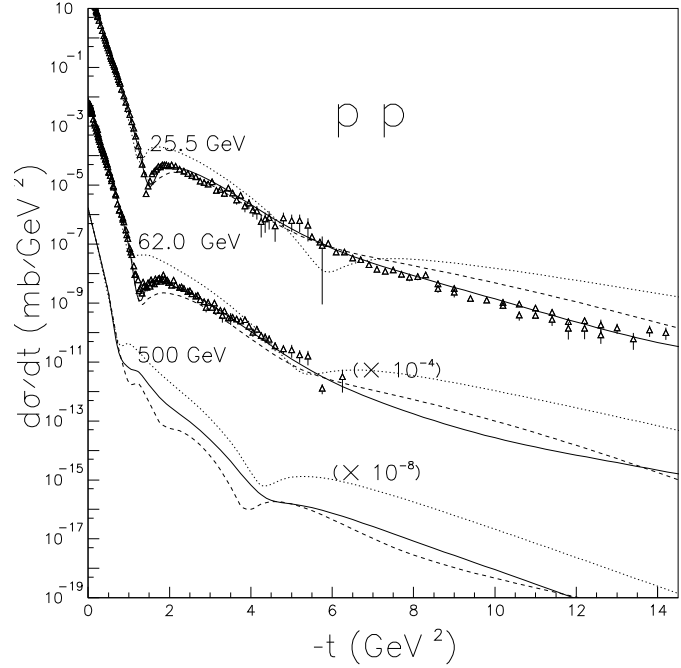}
\end{center}
\vskip -0.5cm
\caption{ Modifications  of a part of the fit in~\cite{poes} and of
prediction at RHIC energy (solid line) when the rescattering
series is approximated with the two-Reggeon exchanges alone (dotted
line) \ie $(n_+,n_-)=(0,0)$, or when the three-Reggeon exchanges are added,
\ie $(n_+,n_-)=(0,0),\ (10),\ (01)$
(dashed line).             }
\end{figure}

\begin{itemize}
\item
Assuming only two-Pomeron exchange as a first approximation is a
pioneering idea to create the dip~\cite{colg} and we know in advance it
should be insufficient. In our language, one extract the Pomeron-Pomeron
contribution from the ten possible two-Reggeon exchanges with
$(n_+,n_-)=(0,0)$. Fig.~7 is an illustrative example of the inadequacy of
this double approximation, for the differential cross-section, as soon as
the transfer is not equal zero.
\item
The next step we consider, is to keep all the ten two-Reggeon exchanges,
\ie when approximating the series (2) with its first term,
$(n_+,n_-)=(0,0)$. The result is shown in Figs.~7-8. The agreement with
the non-truncated series is far from being good. We note a slight
improvement with respect to the two-Pomeron approximation, especially in
reproducing a larger part of the first cone, in creating a dip (although
too high, but unfortunately a second dip is also created) and in reducing
the differential cross-section at high \t\ (which remains still too high).
\item
We go on further by taking into account in addition all the three-Reggeon
exchanges, \ie limiting the series to the first three terms
$(n_+,n_-)=(0,0)$ (two-Reggeon exchanges) and $(n_+,n_-)=(0,1),(1,0)$
(three-Reggeon exchanges). The agreement is better than in the preceding
approximation (see Fig.~8) but is still unsatisfying. Here we can only say
that all the three-Reggeon exchanges contribute significantly to improve
the agreement with the known data and, when these are lacking, to get near
the GE results.
\item
We are in perfect agreement with the statements of the preceding section
concerning the number and specificity of exchanges that must be retained
to approximate the series with its limit, given here by the generalized
eikonalization procedure. From extrapolated differential cross-section
calculations, we find that the upper bound of the index $n_+$ (let us call
it $N_+$) and, to a lesser extend $N_-$, upper bound of $n_-$, leading to
a convergence towards the GE cross-section, increases with $s$ and mostly
with $t$ as indicated in Table 2, giving rise to a very complex picture of
rescatterings in terms of possible diagrams. Whether or not this turns out
to be true, only new data can decide of the quality of the extrapolation
and its consequences.

\end{itemize}

\begin {center}
\begin{tabular}[ht] {|c|c|c||c|c|}
\hline
           &Energy&Transfer &$N_+$ &   $N_-$      \\
           &(GeV) &(GeV$^2$)&      &               \\
\hline
\hline
   RHIC    &200   &   0.0   &   1  &      0     \\
     "     & (or 500)    &   -0.12   &  1   &      0     \\
     "     & "   &   -2.8   &  3   &      1     \\
     "     & "    &   -6.0   &  3   &      1     \\
\hline
      TEVATRON  &2000  &   0.0   &  1   &      0     \\
\hline
   LHC     &14000 &   0.0   &  2   &      0     \\
     "     & "    &   -10.   &  5   &      2     \\
\hline

\end{tabular}
\end {center}
\noindent Table 2 . Upper indexes of the rescattering series $(N_+,N_-)$,
necessary to approximate, within a precision of 1\%, the GE value of the
$pp$ differential cross-section (from the limit of non-truncated series),
extrapolated for the energy and transfer of the future
experiments~\cite{futexp}.

\smallskip
Turning to an other type of approximation, we mention, as a widespread
opinion, that one can neglect the $f-$ and the $\omega-$ contributions
(and consequently, their rescattering terms) at high energy. We have
tested this assertion (giving a sense to the adjective "high"), and found
that the $\omega-$Reggeon (with all its rescatterings) is fully negligible
only if $\sqrt{s}\gsim$ 500 GeV, while for the $f-$Reggeon, the same is
true only at an energy higher than a few TeV.

Clearly, limiting the rescatterings to the three- or two-Reggeon exchanges
(and {\it a fortiori} to the two-Pomeron exchanges) gives a very crude
(wrong) estimation of the angular distribution as soon as $-t$ exceeds
zero. A correct picture of the rescatterings, compatible with presently
available data, requires to limit the series to at least
$(N_+,N_-)=(2,1)$. Indeed, these six couples of indexes correspond to a
quite large number of exchanges (ten for the (0,0) two-Reggeon exchanges,
twenty for the (0,1),(1,0) three-Reggeon exchanges, \etc but we remark
that the complete list of the all four- and five-Reggeon exchanges is not
required).

Finally it is worth pointing out that the substance of the present paper
would have been unchanged if we have illustrated it with the GE Monopole
Pomeron model instead of the GE Dipole Pomeron model.
\bigskip

{\large {\bf Acknowledgements}} We thank L. Jenkovszky for a critical
discussion.

\bigskip
\bigskip

\centerline{\bf APPENDIX}

\medskip\noi
Selecting and completing the information given in~\cite{poes,eikge}, we
collect here the useful formula to understand and perform the rescattering
calculations in the "Generalized Eikonalization" (GE) procedure for the
"Dipole Pomeron"(DP) model. We emphasize that all the relevant expressions
are analytical (they do not require any numerical integration).

\smallskip
{\bf A. Input Born in the $s,t$ space. The Dipole Pomeron model}

\smallskip\noi
We focus on the (dimensionless) Born crossing-even and -odd amplitudes
$a_\pm(s,t)$ of the $pp$ and $\bar pp$ reactions
\footnote{Once again, $+(-)$ correspond to $\bar pp\, (pp)$
process; note that the normalization of~\cite{poes} is used and that the
coupling constants $a_{P}, a_{O}, a_{f}, a_{\omega}$ are reals.}
$$
a_{pp;Born}^{\bar pp}(s,t)=a_+(s,t)\ \pm a_-(s,t)\ ,
\eqno(A1)
$$
starting point to get the eikonalized amplitudes $A(s,t)=A^{\bar pp}_{pp}
(s,t)$ used to fit~:

\noi
i) the total cross-sections
$$
\sigma_{\rm tot} = {4\pi\over s}\Im{\rm m} A(s,t=0) \ ,
\eqno(A2)
$$
ii) the differential cross-sections
$$
{d\sigma \over dt}={\pi\over s^2}\big|A(s,t)\big|^2 \ ,
\eqno(A3)
$$
iii) and the ratios of the real to the imaginary forward amplitudes
$$
\rho={\Re{\rm e} A(s,t=0)\over \Im{\rm m} A(s,t=0)} \ .
\eqno(A4)
$$
The crossing even part in the Born amplitude is a Pomeron, to which a
$f-$Reggeon is added,
while the crossing odd part is an Odderon
(plus an $\omega-$Reggeon)
$$
a_+(s,t)=\ a_{P} (s,t)+\ a_f(s,t) \ ,\quad
a_-(s,t)=\ a_{O} (s,t)+\ a_\omega(s,t) \ .
\eqno(A5)
$$

\noi
For simplicity, in our DP model, the two Reggeons have been taken in the
standard form
$$
a_R(s,t)= a_R\eta_R\tilde s^{\alpha_R(t)}\ e^{b_Rt} ,
\quad (R=f\, {\rm and}\, \omega) ,\quad \eta_f=1,\ \eta_\omega=i\ ,
\eqno(A6)
$$
with linear trajectories
$$
\alpha_R(t)=\alpha_R(0) + \alpha_R'\, t ,
\quad (R=f\, {\rm and}\, \omega) \ .
\eqno(A7)
$$

\noi
Here a "dipole" is chosen for the Pomeron (\ie a linear
combination of a simple pole with a double pole)
$$
a_P(s,t)= a_P^{(D)}(s,t)= a_P \ \tilde s^{\alpha_P(t)}
\left[e^{b_P(\alpha_P(t)-1)} (b_P+\ell n{\tilde s})
                                     \ +\ d_P\ell n{\tilde s}\right]\ .
\eqno(A8)
$$
The Odderon is obtained with the same requirements as for the Pomeron, but
multiplied by a convenient damping factor killing it at $t=0$ in order to
respect the common knowledge
$$
a_O(s,t)=(1-\exp{\gamma t})\ i \  a_O^{(D)}(s,t)\ ;
\eqno(A9)
$$
\ie the amplitude on the r.h.s. $a_O^{(D)}(s,t)$ is constructed along the
same lines as $a_P^{(D)}(s,t)$, changing only the parameters. As usual,
$$
\tilde s\ =\ {s\over s_0} \ e^{-i{\pi\over 2}}\ ,\quad
(s_0=1\ {\rm GeV}^2)\ ,
\eqno(A10)
$$
enforces $s-u$ crossing and $\alpha_i(t)$ are the trajectories
taken, for simplicity, of the linear form
$$
\alpha_i(t)= 1+\delta_i+\alpha_i't  \  ,(i=P,O)\ ,
\eqno(A11)
$$
and verifying the unitarity constraints
$$
\delta_P\ge\delta_O\ , \quad {\rm and}\quad\alpha'_P\ge\alpha'_O\ \ .
\eqno(A12)
$$

\smallskip
{\bf B. Born amplitude in the $s,b$ space}

\smallskip\noi
In eikonal models, the scattering amplitudes are expressed in the
impact-parameter ("$b$") representation ($s,b$ space). First, one defines
the Fourier-Bessel's (F-B) transform
of the Born amplitude
$$
h^{\bar pp}_{pp} (s,b)= {1\over 2s}\int_0^\infty
a^{\bar pp}_{pp;Born} (s,-q^2) J_0(bq) q\, dq
\quad {\rm with} \quad            q=\sqrt{-t}\ .
\eqno(B1)
$$
This is related to the eikonal function ("eikonal" for brevity) by
$$
\chi^{\bar pp}_{pp} (s,b)\  =\ 2\ h^{\bar pp}_{pp} (s,b)\ .
\eqno(B2)
$$
In all eikonalization procedures, one first derives the eikonalized
amplitude in the $b$-representation $H^{\bar pp}_{pp} (s, b)$; the inverse
F-B transform leads then to the usual eikonalized amplitude in the $s,t$
space
$$
A^{\bar pp}_{pp} (s,t)= 2s\ \int_0^\infty
H^{\bar pp}_{pp} (s,b) J_0(b\sqrt{-t}) b\, db\ .
\eqno(B3)
$$
The main technical problem of eikonalization is the
derivation of $H^{\bar pp}_{pp} (s, b)$ once $h^{\bar pp}_{pp} (s,
b)$ are given (for details, see~\cite{eikge}).

\smallskip
Although not required in practice since it is integrated when eikonalizing,
and consequently it is an intermediate quantity,
we give here the expression of the Born amplitude in the impact
parameter representation (half of the eikonal function)
$$
  h^{\bar pp}_{pp} (s,b)\ = \frac{1}{2}\chi^{\bar pp}_{pp}(s,b)\ =
\ h_f(s,b)+h_P(s,b)\ \pm\ \left[h_O(s,b)+h_\omega(s,b)\right]\ ,
\eqno(B4)
$$
with the DP model, defined above.

\noi For the secondary Reggeons, we obtain
$$
  h_{R}(s,b)={1\over2}\eta_{R}a_R{\tilde s^{\alpha_R(0)}\over s}\
{\exp({-{b^2}\over 4B_R})\over 2B_R}=\frac{1}{2}\chi_{R}(s,b),
\eqno(B5)
$$
with (see also $(A6-7)$)
$$
B_R\equiv B_R(s)=
\alpha'_R \ell n\tilde s+b_R \ , \quad R=(f,\omega )\ ,
\quad \eta_f=1,\ \eta_\omega=i\ .
\eqno(B6)
$$
The Pomeron dipole splits into 2 components
$$
  \begin{array}{lll}
    h_{P}(s,b) & = & \displaystyle {-i\ a_P\over 4 \alpha'_P s_0}\ \big(
e^{r_{1,P}\delta_P -{{b^2}\over 4B_{1,P}}}\ +\ d_P\ e^{r_{2,P}
\delta_P -{{b^2}\over 4B_{2,P}}} \big) \\
     & \equiv & \frac{1}{2}\chi_{P}(s,b)\ =\ \frac{1}{2}
     \bigg (\chi_{P1}(s,b)+\chi_{P2}(s,b)\bigg ).
  \end{array}
\eqno(B7)
$$
The Odderon dipole with its damping factor yields
$$
  \begin{array}{lll}
   h_{O}(s,b)&=&\displaystyle { a_O\over 4 \alpha'_{O}s_0}\ \Bigg (
e^{r_{1,O}\delta_O -{{b^2}\over 4B_{1,O}}}\ + d_O\ e^{r_{2,O}
\delta_O -{{b^2}\over 4B_{2,O}}}\Bigg ) \\
 & - & \displaystyle {a_O\over 4 \alpha'_{O}s_0}
 \Bigg (e^{r_{1,O}\delta_O -{{b^2}\over
4\widetilde{B_{1,O}}}}{B_{1,O}\over \widetilde{B_{1,O}}} + d_O\
e^{r_{2,O} \delta_O -{{b^2}\over
4\widetilde{B_{2,O}}}}{B_{2,O}\over \widetilde{B_{2,O}}} \Bigg )\\
& \equiv & \frac{1}{2}\chi_{O}(s,b)=\frac{1}{2}\bigg (
\chi_{O1}(s,b)+\chi_{O2}(s,b)+\widetilde{\chi_{O1}}(s,b)+
\widetilde{\chi_{O}}(s,b)\bigg).
 \end{array}
\eqno(B8)
$$
We have defined (see also $(A8-11)$)
$$
 r_{1,J}\equiv r_{1,J}(s)=\ell n\tilde s+b_J\ ,
 r_{2,J}\equiv r_{2,J}(s)=\ell n\tilde s\ , (J=P,O)\ ,
\eqno(B9)
$$
and
$$
\begin{array}{lll}
B_{i,P}&\equiv& B_{i,P}(s)=\alpha'_P r_{i,P} \ ,
\quad B_{i,O}\equiv B_{i,O}(s)= \alpha'_O r_{i,O} \ ,\\
\widetilde{B_{i,O}}&\equiv&\widetilde{B_{i,O}}(s)
= \alpha'_O r_{i,O} +\gamma \ , (i=1,2)\ .
\end{array}
\eqno(B10)
$$

\smallskip
{\bf C. Rescattering Series}

\smallskip\noi
We rewrite the rescattering series (2)
(part of the eikonalized amplitude added to the Born contribution (1))
$$
A^{\bar pp}_{pp,rescat}(s,t)\ =\ \sum_{n_+=0}^\infty \sum_{n_-=0}^\infty
  a^{\bar pp}_{pp;n_+,n_-}(s,t) \ ,
\eqno(C1)
$$
with the ($n_+,n_-$) term in the general case of three parameters
($\lambda_\pm, \lambda_0$) of the GE procedure
$$
\begin{array}{lll}
  a^{\bar pp}_{pp;n_+,n_-}(s,t) & = &
\ i\ s\ {(i\lambda_+)^{n_+}\ (\pm i\lambda_-)^{n_-} \over
(n_++n_-+2)! } \\
& \times & \left(F_{n_+,n_-}(z)\cdot
\displaystyle{I}+F_{n_-,n_+}(z)\cdot \displaystyle{II}+
G_{n_+,n_-}(z)\cdot \displaystyle{III}\right) \ ,
\end{array}
\eqno(C2)
$$
where the hypergeometric function $_{2}F_{1}$ has been introduced
in $F$ and $G$, functions of the argument
$z=\frac{\lambda_{0}^{2}}{\lambda_{+}\lambda_{-}}$
$$
  \begin{array}{lll}
    F_{n_\pm,n_\mp}(z) & = & z (n_\pm+1)\cdot _2
    F_1(1-n_\mp,-n_\pm;2;z)\cdot
(1-\delta_{n_\mp,0}) + \delta_{n_\mp,0} \ ,\\
G_{n_+,n_-}(z) &= &_2 F_1 (-n_-,-n_+;1;z)\ .
  \end{array}
\eqno(C3)
$$
The inverse Fourier-Bessel transforms are
the following three functions in the $s,t$ space~:
$$
     I  = \lambda_+ \sum _{\ell =0}^{n_++2} \sum_{m=0}^{n_-}
\pmatrix{n_++2\cr \ell\cr}\pmatrix{n_-\cr m\cr}\cdot{\rm Int}\
_{n_++2-\ell, n_--m,\ell ,m} (s,t)\ , \eqno(C4)
$$
$$
 II = \lambda_- \sum
_{\ell =0}^{n_+}\sum_{m=0}^{n_-+2} \pmatrix{n_+\cr
\ell\cr}\pmatrix{n_-+2\cr m\cr}\cdot{\rm Int}\ _{n_+-\ell, n_-+2-m,\ell
,m} (s,t)\ , \eqno(C5)
$$
$$
 III  =\ \pm 2{\lambda_+\lambda_-\over\lambda_0}
\sum _{\ell =0}^{n_++1}\sum_{m=0}^{n_-+1} \pmatrix{n_++1\cr
\ell\cr}\pmatrix{n_-+1\cr m\cr}\cdot{\rm Int}\ _{n_++1-\ell, n_-+1-m,\ell
,m} (s,t)\ . \eqno(C6)
$$
$\ds{{n \choose k}}$ is the binomial coefficient and ${\rm
Int}(s,t)$ is the integral over the four eikonals defined above \ie
$$
  {\rm Int}\ _{\lambda,\mu,\ell,m}\ (s,t)=\int_0^\infty
\chi_P^\lambda(s,b)\,\chi_O^\mu(s,b)\,\chi_f^\ell(s,b)\,
\chi_{\omega}^m(s,b)\; J_0(b\sqrt{-t})\, b\, db.
\eqno(C7)
$$
When the dipole Odderon does contain a damping factor at
$t=0$, this integral writes
$$
  \begin{array}{ll}
  &{\rm Int}_{\lambda,\mu,\ell,m} (s,t)=
  C\ \ds{\sum \limits_{\lambda'=0}^{\lambda}
     \sum \limits_{\sigma=0}^{\mu}\sum \limits_{\mu'=0}^{\mu-\sigma}
     \sum \limits_{\nu=0}^{\sigma} }
    \pmatrix{\lambda\cr\lambda'\cr}\pmatrix{\mu\cr\sigma\cr}
    \pmatrix{\mu-\sigma\cr\mu'\cr}\pmatrix{\sigma\cr\nu\cr} \\
    & \times\ \exp\bigg[\ds{r_{1,P}\delta_{P}(\lambda-\lambda')+
    r_{2,P}\delta_{P}\lambda'+r_{1,O}\delta_{O}(\mu-\mu'-\nu)+
    r_{2,O}\delta_{O}(\mu' +\nu)}\bigg] \\
    & \times\ d_{P}^{\lambda'}d_{O}^{\mu '+\nu}
    \Bigg(\ds -\frac{B_{1,O}}{\widetilde{B_{1,O}}}\Bigg)^{\sigma -\nu}
    \Bigg(\ds -\frac{B_{2,O}}{\widetilde{B_{2,O}}}\Bigg)^{\nu}\ \cdot \
    {int(s,t)} \ ,
  \end{array}
\eqno(C8)
$$
where
$$
  C \equiv C(s) =
  \left (\frac{-ia_{P}}{2\alpha'_{P}s_{0}}\right)^{\lambda}
  \left (\frac{a_{O}}{2\alpha'_{O}s_{0}}\right)^{\mu}
  \left (\frac{a_{f}\tilde s^{\alpha_{f}(0)}}{2B_{f}s}\right)^{\ell}
  \left (\frac{ia_{\omega}\tilde s^{\alpha_{\omega}(0)}}
  {2B_{\omega }s}\right)^{m};
\eqno(C9)
$$
$$
   int(s,t)\ =\ \int_{0}^{\infty}\ \exp\bigg( \frac{-Db^{2}}{4}\bigg)
   J_{0}(b\sqrt{-t})bdb=\frac{2}{D}\exp\bigg( \frac{t}{D}\bigg);
\eqno(C10)
$$
$$
D \equiv D(s)=\frac{\lambda -\lambda'}{B_{1,P}}+
      \frac{\lambda'}{B_{2,P}}+
      \frac{\mu -\sigma -\mu'}{B_{1,O}}+
      \frac{\mu'}{B_{2,O}}+
      \frac{\sigma -\nu}{\widetilde{B_{1,O}}}+
      \frac{\nu}{\widetilde{B_{2,O}}}+
      \frac{\ell}{B_{f}}+\frac{m}{B_{\omega}}\ .
\eqno(C11)
$$

\bigskip

\bigskip
\bigskip

\end{document}